\documentclass[showpacs, prb, twocolumn]{revtex4}
\usepackage{latexsym}
\usepackage{amsfonts}
\usepackage{amssymb}
\usepackage{amsmath}
\usepackage{graphicx}
\usepackage{bm}
\begin{document}

%\preprint{APS/123-QED}

\title{Electrical detection of spin pumping: dc voltage generated by ferromagnetic resonance at ferromagnet/nonmagnet contact}

\author{M. V. Costache$^{1,2}$}
%\altaffiliation[Present address: ]{Francis Bitter Laboratory, Massachusetts Institute of Technology, Cambridge, MA 02139,USA. \\
%E-mail: costache@mit.edu}
%\altaffiliation{E-mail: costache@mit.edu}
%Present address: California Institute of Technology,
%Pasadena, CA 91125, USA.

\author{S. M. Watts$^1$, C. H. van der Wal$^1$}

\author{B. J. van Wees$^1$}
\affiliation{$^{1}$Physics of Nanodevices Group, University of Groningen, Nijenborgh 4, 9747 AG Groningen, The Netherlands.\\
$^{2}$Massachusetts Institute of Technology, Cambridge, MA 02139,USA. }

\date{\today}

\begin{abstract}

We describe electrical detection of spin pumping in metallic nanostructures. In the spin pumping effect, a precessing ferromagnet attached to a normal-metal acts as a pump of spin-polarized current, giving rise to a spin accumulation. The resulting spin accumulation induces a backflow of spin current into the ferromagnet and generates a dc voltage due to the spin dependent conductivities of the ferromagnet. The magnitude of such voltage is proportional to the spin-relaxation properties of the normal-metal. By using platinum as a contact material we observe, in agreement with theory, that the voltage is significantly reduced as compared to the case when aluminum was used.
\\Furtheremore, the effects of rectification between the circulating rf currents and the magnetization precession of the ferromagnet are examined. Most significantly, we show that using an improved layout device geometry these effects can be minimized.
\end{abstract}


\maketitle

\section{Introduction}

During the last several years there has been a continuing interest in high frequency phenomena in spintronic devices, as they are expected both to provide applications for microwave signal-processing, and to become a powerful new tool for fundamental studies of spin dynamics in magnetic nanostructures \cite{slons1,berger01,Kiselev,Tulapurkar,sankey}.

It was predicted by Slonczewski \cite{slons1} and Berger \cite{berger01} that angular momentum is transferred from spin polarized currents to the magnetization of the ferromagnets when charge currents are sent trough spin valves with non-collinear magnetizations (i.e. spin torque effect). This can excite and even switch the magnetization direction of the softer ferromagnet. Experiments with pillar-type structures \cite{katine,grollier,sun1} confirmed these predictions.

It is natural to expect that if a spin current can induce magnetization motion the reciprocal process may also be possible: a moving magnetization in a ferromagnet can emit a spin current into an adjacent conductor. This effect is the so-called spin pumping, proposed by \citet{bauer02a} and \citet{brataas}. Spin pumping is a mechanism where a pure spin current, which does not involve net charge currents, is emitted at the interface between a ferromagnet with a precessing magnetization and a normal-metal region. It is an important mechanism to generate spin currents, since other electronic methods based on driving an electrical current through a ferromagnet/semiconductor interface are strongly limited by the so-called conductance mismatch \cite{schmidt:2000}. Berger \cite{berger02} proposed a similar mechanism to generate a dc voltage by ferromagnetic resonance (FMR), based on spin-flip scattering in the ferromagnet as induced by spin waves.

Recently, spin pumping has been demonstrated in ferromagnetic resonance experiments with thin multilayers, where it appears as an enhancement of the Gilbert damping constant of magnetization dynamics \cite{Mizukami02,Urban01,Heinrich03,Lenz,Imamura07}, and using time resolved magneto-optic Kerr effect \cite{woltersdorf:246603}. Although these experiments are very important in providing evidence for the spin pumping mechanism, the detection technique can be viewed as an indirect method to measure the spin pumping effect. Several experimental methods have been proposed to electrically detect the spin pumping mechanism \cite{brasil05,Saitoh_apl}. The general problem of these methods is the rectification effects at the ferromagnet/normal-metal contact, which can suppress or mimic the spin pumping signal \cite{meAMR,gui,Yamaguchi}. Thus, the identification of these spurious effects is crucial and represents one of the main themes of this paper.

In a recent paper, \cite{mePRL} we have demonstrated spin pumping with a single permalloy strip in an electronic device, in which it is directly detected as a dc voltage signal. In this paper, we describe additional experiments on spin pumping effect, designed explicitly to eliminate the rectification effects. We explain in more detail the theoretical prediction for the voltage, and we identify and quantify different contributions of the rectification effect. Importantly, we show that by using an appropriate device geometry, these effects can be minimized.

\section{Spin pumping effect}\label{pumping_ch2}
\begin{figure}[h]
\centering
\includegraphics[width = 8.6cm]{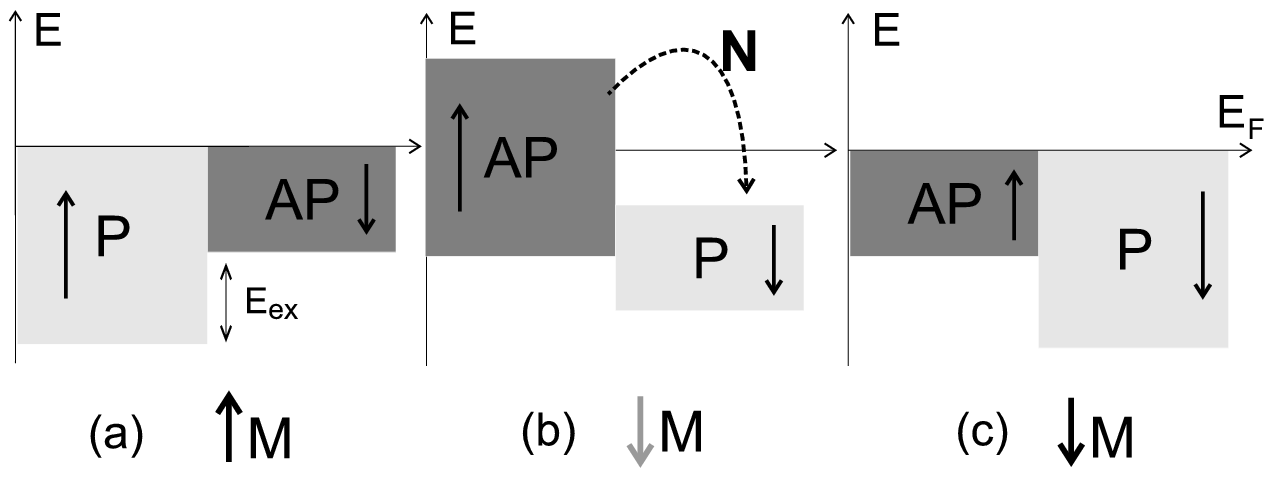}
\caption{Simplified picture of the spin pumping process. (a) Population of spin-up and spin-down bands in equilibrium. (b) Situation after sudden reversal of the magnetization direction. The arrows denote spin flow from one spin population to another one. (c) Equilibrium situation again but with magnetization in opposite direction. [Adapted from ref. \cite{brataas_talk}]} \label{pic_pump}
\end{figure}
As discussed above, the emission of a spin current into a conductor by a moving magnetization of an adjacent ferromagnet is essentially the reciprocal of the spin torque mechanism in spin valves, where the magnetization is excited by a spin current.

A simplified picture of the process is schematically shown in Fig. \ref{pic_pump}. We consider a F/N junction at equilibrium, where in F exist a larger population of spins in the direction of magnetization, than antiparallel. When the magnetization direction is suddenly switched, the bands instantaneously shift in energy. However, in order to go back to the equilibrium situation there has to be spin transfer from one spin population to another (i.e. spin relaxation). If F is in contact with N, this transfer of spins can go via N. Thus the spin relaxation process for F is modified when it is in contact with an adjacent N, and depends on the spin relaxation properties on N. As a result, an ac spin current is emitted into N when the magnetization is switched back and forth under an oscillating magnetic field. \citet{bauer02a} analyzed the case of circular precession of the magnetization and found that in addition to the ac current, a dc spin current is also emitted. A way to periodically change the magnetization direction is to put F into FMR, where circular precession of the magnetization can be resonantly excited by a small applied rf magnetic field \cite{meACAMR,meAMR}.

The transfer of spin angular momentum by the precessing magnetization of F in contact with N (spin pumping) was first described \cite{bauer02a} using the formalism of parametric charge pumping \cite{Büttiker} developed in the context of mesoscopic scattering problems. The main points of this description are discussed below.

\begin{figure}[h]
\centering
\includegraphics[width = 8.5cm]{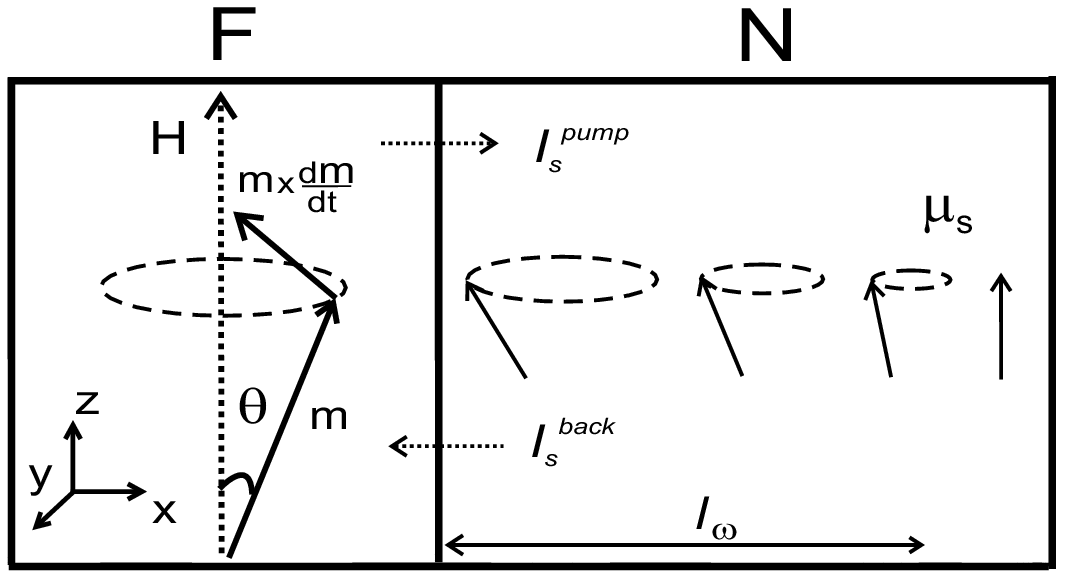}
\caption{The F/N structure in which the resonant precession of the magnetization direction $\mathbf{m}$ pumps a spin current $\bm{I}^{pump}_{s}$ into N. The spin pumping builds up a spin accumulation $\bm{\mu}_{s}^{N}$ in N that drives a spin current $\bm{I}^{back}_{s}$ back into the F. The component of the $\bm{I}^{back}_{s}$ parallel to \textbf{m} can enter into F. Since the interface and the bulk conductances of F are spin dependent, this can result in a dc voltage across the interface.} \label{pump_ch2}
\end{figure}

\textbf{Spin currents at the interface.} As illustrated in Fig. \ref{pump_ch2}, a spin current $\mathbf{I}_{s}^{pump}$ is pumped by the (resonant) precession of a ferromagnet magnetization into an adjacent normal-metal region. Assuming the F at FMR state, \citet{bauer02a} have calculated the spin pumped current using a scattering matrix approach based on the microscopic details of the interface,
\begin{equation}
\bm{I}_{s}^{pump}=\frac{\hbar}{4\pi} g^{\uparrow\downarrow} \mathbf{m}\times
\frac{d\mathbf{m}}{dt}\;. \label{pump_crch2}
\end{equation}
where \textbf{m} represents the magnetization direction. $g_{\uparrow\downarrow}$ is the real part of the mixing conductance \cite{PhysRevB.62.5700,bauer00}, a material parameter which describes the transport of spins that are noncollinear to the magnetization direction at the interface and is proportional to the torque acting on the ferromagnet in the presence of a noncollinear spin accumulation in the normal metal \cite{PhysRevB.66.014407,PhysRevB.65.220401}. This equation shows that the spin current, which goes into N, is perpendicular both to the magnetization direction \textbf{m} and to the change of \textbf{m} in time. This current has ac and dc components, but in the limit $\omega\tau_{N}~\gg$ 1 (see later discussion), the time-averaged pumping current reads \cite{brataas} $|\langle{\bm{I}}_{s}^{pump}\rangle_{t}|=I_{dc}=\hbar\omega
g^{\uparrow\downarrow}sin^{2}\theta/4 \pi$.

Depending on the spin related properties of the N, the spin current emission has two limiting regimes. When the N is a good ``spin sink" (in which spins relax fast), the injected spin current is quickly dissipated and this corresponds to a loss of angular momentum and an increase in the effective Gilbert damping of the magnetization precession. This has been observed experimentally in nano-pillar structures \cite{Mizukami02,Urban01,Heinrich03,Lenz,Imamura07}. The total spin current is given by $\bm{I}_{s}^{pump}$.

The opposite regime is when the spin-flip relaxation rate is smaller than the spin injection rate. In this case, a spin accumulation $\bm{\mu}_{s}$ builds up in the normal metal (Fig. \ref{pump_ch2}). The spin accumulation can diffuse away from the interface, but can also diffuse back into the F. This back flow current is given by
\begin{equation}
\bm{I}_{s}^{back}  =  \frac{g_{\uparrow \downarrow }}{2\pi
N}[\bm{\mu}_{s}-\mathbf{m}(\mathbf{m} \cdot \bm{\mu}_{s})]\;,
\end{equation}
where $N$ is the one-spin density of states. The total spin current in this case is
$\bm{I}^{F}_{s}=\bm{I}_{s}^{pump}+\bm{I}_{s}^{back}$.

\textbf{Spin battery.} A spin battery operated by FMR has been proposed by the \citet{brataas} in the limit of weak spin-flip scattering in the F. The spin accumulation in N can be calculated by solving the spin diffusion equation
\begin{equation}
\frac{\partial \bm{\mu}_{s}}{\partial t}=D_{N}\frac{\partial ^{2}\bm{\mu}_{s}}{
\partial x^{2}}-\frac{\bm{\mu}_{s}}{\tau_{N}}\label{diffuspump1}
\end{equation}
where $\tau_{N}$ is the spin-flip time and $D_N$ is the diffusion coefficient in N.
We assume that the spin diffusion length in N is much larger than the spin precession length $l_{\omega}\equiv\sqrt{D_N/\omega}$ ($\omega$ is precessional frequency), i.e.
$\lambda_{N}=\sqrt{D_{N}\tau_{N}} \gg l_{\omega}$, or equivalent by $\omega\tau_{N}~\gg$ 1. This means that if the length of N is larger than $l_{\omega}$, the x, y components of spin accumulation are fully averaged (due to dephasing) and the remaining z component is constant and along the static magnetic field direction \cite{brataas}. The time-averaged spin accumulation  $\langle\bm{\mu}_{s}\rangle _{t}=\mu_{z}$\textbf{z} in the N close to the interface reads\cite{brataas}
\begin{equation}
\mu_{0,z} =\hbar \omega \frac{sin^{2}\theta}{sin^{2}\theta + \eta}\;,
\label{eq_pump_ch2}
\end{equation}
where $\theta$ is the precession cone angle and $\eta$ is a reduction factor determined by the ratio between injection time and spin-flip relaxation time.

Intuitively, the spin accumulation can be measured electrically using a second ferromagnet as a spin dependent contact, placed at a shorter distance compared to the spin-flip length \cite{jedema:Nat2002,Otani,valenzuela01,brataas}.

\textbf{Voltage at F/N interface.} Importantly, \citet{Xuhui} have predicted a more direct way to detect the spin pump effect in which the precessing ferromagnet acts also as the detector. We have to take into account that the spin accumulation $\bm{\mu}_{s}$ in a diffusive metal drives the spin current $\bm{I}_{s}^{back}$ back into the F. The component parallel to \textbf{m} can enter F. Moreover, since the interface and the bulk conductances of F are spin dependent, this can result in charge accumulation, close to the interface, and thereby a dc voltage across the interface. The
chemical potential difference across the interface has been calculated by \citet{Xuhui} following the lines of the \citet{brataas} model, but including the spin diffusion back into F and spin-relaxation in F.
As mentioned above, the relevant length-scale for the averaging of the transverse (x, y) components of the spin current is $l_{\omega}$. Therefore for a device with dimensions larger than $l_{\omega}$ the spin-up (down) effective conductances $g_{\omega}^{\uparrow(\downarrow)}$ of the interface are composed of the interface conductances $g^{\uparrow(\downarrow)}$ in series with a conductance of the bulk N over a length scale of $l_{\omega}$. These relations are given by
$g_{\omega}^{\uparrow(\downarrow)}=g^{\uparrow(\downarrow)}/(1+ g^{\uparrow(\downarrow)}/g_{\omega})$ and the mixing conductance
$g_{\omega}^{\uparrow\downarrow}=g^{\uparrow\downarrow}/(1+g^{\uparrow\downarrow}/g_{\omega})$,
where $g_{\omega}=(\sigma_{N} A)/ l_{\omega}$ (A is the area of the interface).
Polarization $p_{\omega}=(g_{\omega}^\uparrow-g_{\omega}^\downarrow)/
(g_{\omega}^\uparrow+g_{\omega}^\downarrow)$ is also introduced.

In the limit of large spin-flip in F and the size of N $\gg \lambda_{N}$ and for small angle precession ($\theta\rightarrow 0$), the chemical potential difference is given by \cite{Xuhui}
\begin{equation}
\Delta \mu_{0}=\frac{p_{\omega}
g_{\omega}^{\uparrow\downarrow}}{2(1+\frac{g_{N}}{g_{F}})(1-p_{\omega}^{2})
(g_{\omega}^\uparrow+g_{\omega}^\downarrow)+2g_{N}}\theta^{2} \hbar \omega \;,
\label{xuhui_ch2}
\end{equation}
where $g_{N}$ ($g_{F}$) is the conductance of the bulk N (F) over a length scale of $\lambda_{N}$ ($\lambda_{F}$). For a thorough review of the above discussion see ref.\cite{Xuhui}.

\textbf{Interface currents matching.} In this section, we describe a simple way to find the voltage (similar to Eq. 5) using spin-current matching at the interface. By writing all the currents involved in the process and matching them at the interface, all components of the spin accumulation at the interface can be determined. It is convenient to transform the equations into a rotating frame of reference in which the uniform magnetization motion can be formally eliminated, and the unit magnetization vector is $\hat{\mathbf{m}}=(sin\theta,0,cos\theta)$. Basically, for this problem we have to consider three currents with their components. First, the spin pumping current (Eq. \ref{pump_crch2}) is given by
\begin{equation}
I_{s , \perp}^{pump}=  g^{\uparrow\downarrow} sin \theta \hbar \omega \;.
\label{pump1_ch2}
\end{equation}
Second, the back flow current consists of components parallel and perpendicular to $\hat{\mathbf{m}}$, and can be written in terms of spin accumulation $\bm{\mu_{0}}$ at the interface,
\begin{eqnarray}
\bm{I}_{s, \parallel}^{back}= g_{F}\bm{\mu}_{0, \parallel}; \nonumber\\
\bm{I}_{s, \perp}^{back} = g^{\uparrow\downarrow}\bm{\mu}_{0, \perp}
\;. \label{back1_ch2}
\end{eqnarray}

The sum of Eqs. \ref{pump1_ch2} and \ref{back1_ch2}, gives the total spin current on the F side of the interface. Third, the spin current on the N side of the interface is found by solving the Bloch equations for the spin accumulation in N, from this the current at the interface is given by \cite{steve_prl}

\begin{equation}
\bm{I}_{s}^{N} =  g_{\omega}\left(
\begin{array}
[c]{c}
\mu_{0,x}-\mu_{0,y}\\
\mu_{0,x}+\mu_{0,y}\\
\frac{g_{N}}{g_{\omega}}\mu_{0,z}
\end{array}
\right)\;,
\end{equation}
in terms of $\bm{\mu}_{0}$ at the interface. This current has three components. The z component is determined only by the usual spin relaxation process. For the x and y components, two effects are important: precession, which results in mixing of the two components, depending on the time spent in N; and averaging, which reduces the total amplitude of the components. The spin accumulation $\bm{\mu}_{0}$ is determined by matching the currents at the interface $\bm{I}_{s}^{N}=\bm{I}_{s}^{F}= \bm{I}_{s}^{back}+\bm{I}_{s}^{pump}$. The dc voltage at the interface is proportional to the projection of $\bm{\mu}_{0}$ onto $\hat{\mathbf{m}}$, and for the limit $g^{\uparrow\downarrow}\geq g_{\omega}$ is given by
\begin{equation}
V = -p \bm{\mu}_{0}\cdot\hat{\mathbf{m}}\simeq -p \frac{g_{\omega}}{g_{F}}\left(
1-\frac{g_{N}}{g_{\omega}}\right)\cos\theta\sin^{2}
\theta\hbar\omega \;. \label{voltage_ch2}
\end{equation}
The simple form of Eq. \ref{voltage_ch2} results from the relative independence of the dc voltage on $g^{\uparrow\downarrow}$ (physical argument of this result remains to be clarified). For our devices (N = Al and F = Py). Using $\sigma_{F}=6.6\cdot10^{6} \Omega^{-1}$m$^{-1}$, $\sigma_{N}=3.1\cdot10^{7} \Omega^{-1}$m$^{-1}$, $\lambda_{F}=5$ nm, $\lambda_{N}=500$ nm and $l_{\omega}=300$ nm we estimate the conductances at room temperature:
\begin{eqnarray}
g_{F}/A  & = &\sigma_{F}/\lambda_{F}\simeq1\cdot10^{15}{\Omega^{-1} m^{-2}}\nonumber\\
g_{\omega}/{A} & = &\sigma_{N}/{l_{\omega}}\simeq1\cdot10^{14}{\Omega^{-1} m^{-2}}\nonumber\\
g_{N}/{A}  & = &\sigma_{N}/{\lambda_{N}}\simeq8\cdot10^{13}{{\Omega^{-1} m^{-2}}}\;.
\end{eqnarray}
And according to \citet{xia1}, $g^{\uparrow\downarrow}/{A}\simeq5\cdot10^{14}\Omega^{-1} $m$^{-2}$.
For a quantitative assessment of the relations \ref{xuhui_ch2} and \ref{voltage_ch2} we assume $\theta\approx5^{\circ}$ ($sin^{2}\theta=0.01$) and $\omega=10^{11}s^{-1}$ ($\hbar\omega= 65~\mu $eV). First Eq. \ref{voltage_ch2}, by using $p=0.4$ we find dc voltage $\approx$ 20 nV. Second Eq. \ref{xuhui_ch2}, by using $p_{\omega}=0.06$, ${g^{\uparrow}}/{A}=0.31\times10^{15}\Omega^{-1} $m$^{-2}$ and ${g^{\downarrow}}/{A}=0.19\times10^{15}\Omega^{-1}$m$^{-2}$ (from ref.  \cite{jedema_thesis}), the dc voltage is of the same order of magnitude $\approx$ 20 nV.

\section{Experimental Procedures}\label{exp_data}

\begin{figure}[h]
\centering
\includegraphics[width=7.5cm]{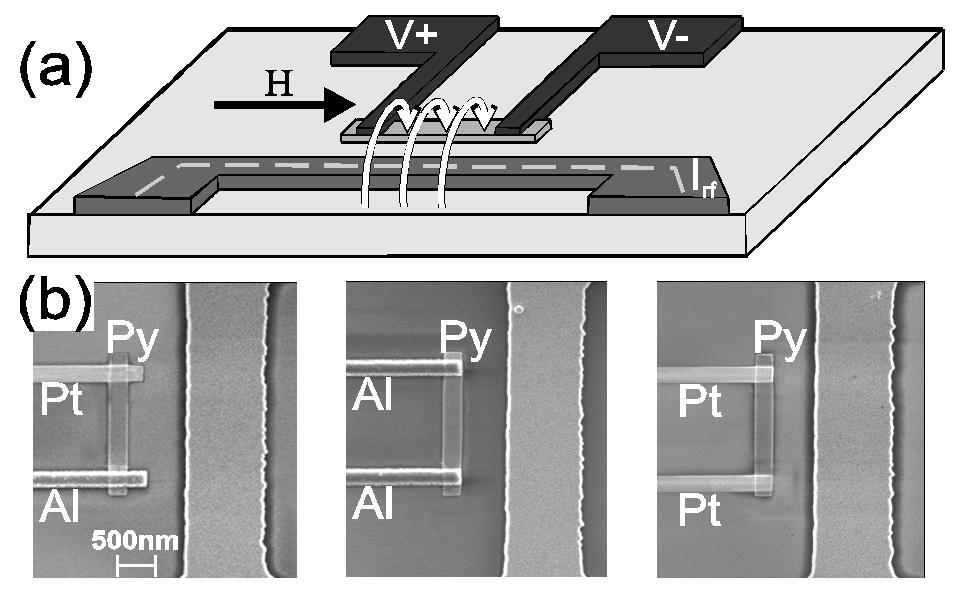}\\
\caption{(a) Schematic diagram of the device. On the lower
side, through the shorted-end of a coplanar strip a current
$I_{rf}$ generates an rf magnetic field, denote by the
arrows. The Py strip in the center produces a dc voltage
$\Delta V=V^{+}-V^{-}$. H denotes the static magnetic field
applied along the strip. (b) Scanning electron microscope
pictures of the central part of the devices.} \label{str}
\end{figure}

Our detection technique is based on the asymmetry in the spin pumping effect between two contacts in a device geometry where a ferromagnet is contacted with two normal-metal electrodes.
The largest such asymmetry is obtained when one of the metal electrodes is a spin sink such as Pt, for which we expect a negligible contribution, while the other has a small spin flip relaxation rate, such as Al. Therefore, we anticipate a net dc voltage across a Py strip contacted by Pt and Al electrodes when the ferromagnet is in resonance.

Additional, we studied control devices where the Py strip is contacted by the same material Pt and Al. For these devices we expected no signal because: (i) The voltages for identical interfaces are the same and their contribution to $\Delta V$ cancels. (ii) Pt has a very short spin diffusion length, resulting in a small spin accumulation, a small backflow and thus a lower signal.
\begin{figure}[ht]
\centering
\includegraphics[width = 6cm]{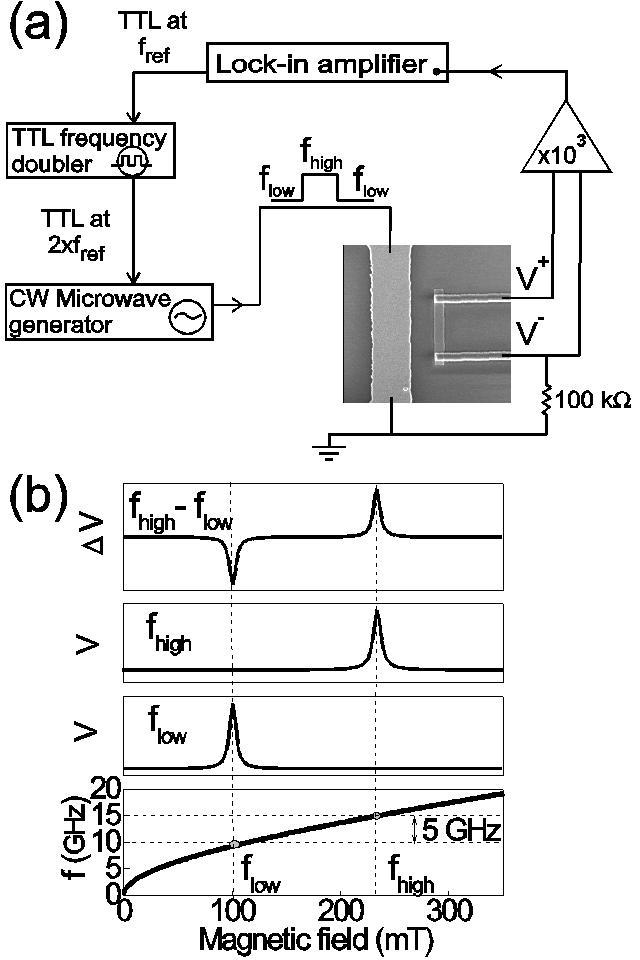}
\caption{Schematic diagrams of the experimental setup and of the microwave frequency modulation method. (a) A TTL signal at a reference frequency $f_{ref}$ (17 Hz) generated by a Lock-in Amplifier (master device) is first fed into a frequency doubler. Then, the TTL at 2x$f_{ref}$ is fed into a CW Microwave Generator. At each TTL input, the CW Generator provides frequency hopping of the rf current switching between $f_{high}$ and $f_{low}$ at $f_{ref}$. The dc voltages produced by the device are amplified and detected by the Lock-in Amplifier as a difference $\Delta V=V(f_{high})-V(f_{low})$. (b) At the
bottom, the resonant frequency dependence on the static magnetic field is shown. Next, the diagrams of the dc voltage vs. static magnetic field corresponding to the resonance at high and low frequencies. On top, the measured difference in dc voltage between the two frequencies, $\Delta V=V(f_{high})-V(f_{low})$ is plotted.} \label{setup}
\end{figure}

Figure \ref{str}(a) shows a schematic illustration of the lateral devices used in the
present study. The central part of the device is a ferromagnetic strip of permalloy (Ni$_{80}$Fe$_{20}$, or Py) connected at both ends to normal metals, Al and/or Pt ($V^{-}$ and $V^{+}$ contacts). The devices are fabricated on a Si/SiO$_{2}$ substrate using e-beam lithography, material deposition and lift-off. A 25 nm thick Py strip with 0.3$\times$3 $\mu m^2$ lateral size was e-beam deposited in a base pressure of 1x$10^{-7}$ mBar. Prior to deposition of the 30 nm thick Al or/and Pt contact layers, the Py surface was cleaned by Ar ion milling, using an acceleration voltage of 500 V with a current of 10 mA for 30 sec, removing the oxide and few nm of Py material to ensure transparent contacts.
We measured in total 17 devices (this includes 4 devices with a modified contact geometry, described later in the paper). The different contact material configurations are shown in Fig. \ref{str}(b).

Figure \ref{setup}(a) illustrates the experimental setup for the measurements. We measured the dc voltage generated between the $V^{+}$, $V^{-}$ electrodes as a function of a slowly sweeping magnetic field ($H$) applied along the Py strip, while applying an rf magnetic field ($h_{rf}$) perpendicular to the strip.

We have recently shown that a submicron Py strip can be driven into the uniform precession ferromagnetic resonance mode \cite{meACAMR,meAMR} by using a small perpendicular rf magnetic field created with an on-chip coplanar strip waveguide \cite{Gupta} (CSW) positioned close to Py strip (similar geometry as shown in Fig. \ref{str}). For the applied rf power 9 dBm, an rf current of $\approx12~mA$ $rms$ passes through the shorted-end of the coplanar strip waveguide and creates an rf magnetic field with an amplitude of $h_{rf}\approx1.6$ mT at the location of the Py strip  \cite{cur}. We confirmed with anisotropic magnetoresistance (AMR) measurements \cite{meAMR} that on-resonance the precession cone angle is $\approx5^{\circ}$.

In order to reduce the background (amplifier) dc offset and noise we adopted a lock-in microwave frequency modulation technique. During a measurement where the static magnetic field is swept from -400 mT to +400 mT, the rf field is periodically switched between two different frequencies and we measured the difference in dc voltage between the two frequencies $\Delta V=V(f_{high})-V(f_{low})$ using a lock-in amplifier. For all the measurements the lock-in frequency is 17 Hz and the difference between the two microwave frequencies is 5 GHz. A diagram of the measurement method is shown in Fig. \ref{setup}(b).

\section{Results and Discussion}
\subsection{Detection of Spin pumping}

\begin{figure}[h]
\centering
\includegraphics[width=7.5cm]{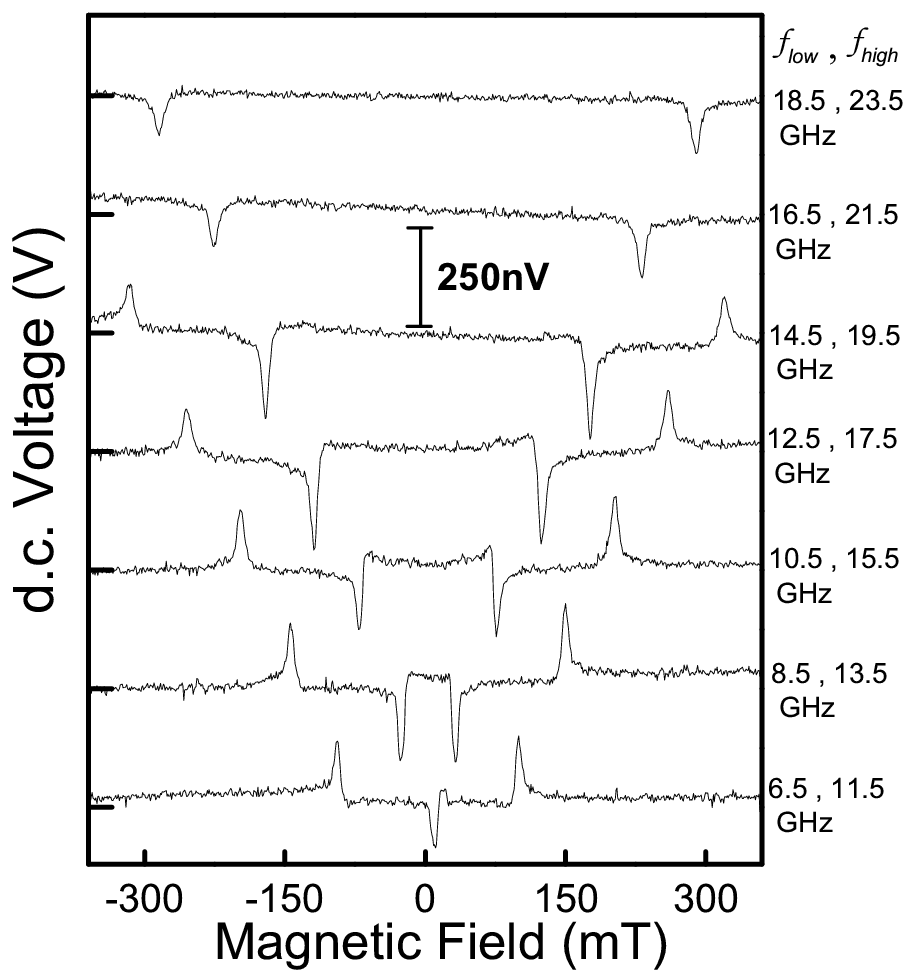}
\caption{The dc voltage $\Delta V$ generated by a Pt/Py/Al device in response to the rf magnetic field plotted as a function of the static magnetic field. The frequencies of the rf field are as shown. The peaks (dips) correspond to resonance at $f_{high} (f_{low})$. The data are offset vertically, for clarity. \cite{mePRL}} \label{fig2}
\end{figure}

Here, we describe precise, room-temperature measurements of the dc voltage across a Py strip contacted by Pt and Al electrodes when the ferromagnet is in resonance. Figure \ref{fig2} shows the electric potential difference $\Delta V$ from a Pt/Py/Al device. Sweeping the static magnetic field in a range -400 mT to +400 mT, a peak and a dip like signal are observed at both positive and negative values of the static field. Since we measured the difference between two frequencies, the peak corresponds to the high resonant frequency ($f_{high}$) and the dip to the low resonant frequency ($f_{low}$), see Fig. \ref{setup}(b). For the opposite sweep direction, the traces are
mirror image. We measured 8 devices with contact material Pt/Py/Al. The measured resonances are all in the range +100 nV to +250 nV. Notably, the dc voltages are all of the same sign (always a peak for $f_{high}$), meaning that for Pt/Py/Al devices, the Al contact at resonance is always more negative than the Pt contact.

\begin{figure}[h]
\centering
\includegraphics[width = 8.5cm]{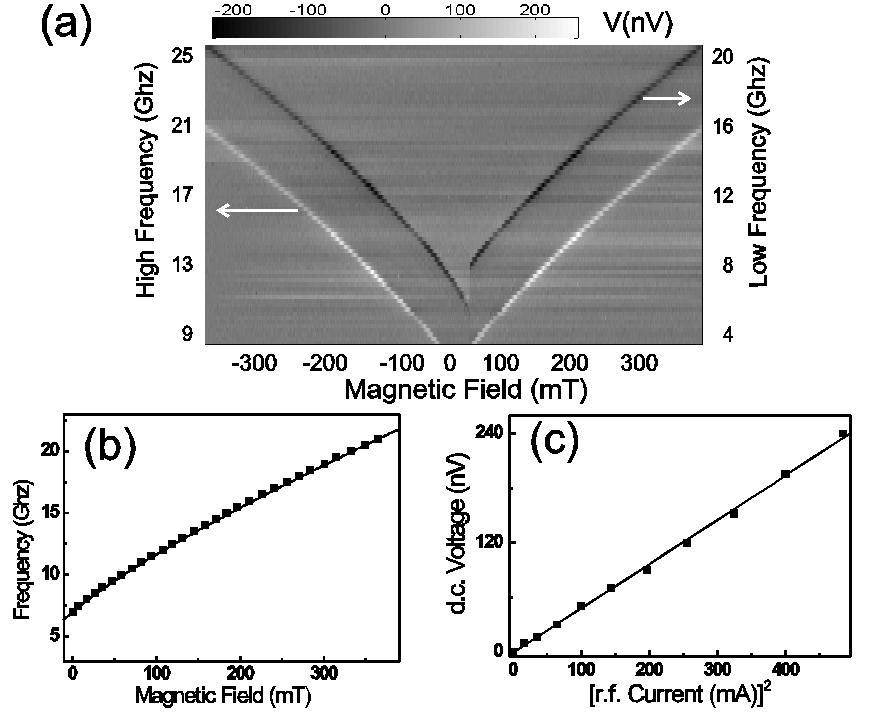}
\caption{(a) Gray scale plot of the dc voltage $\Delta V$,
measured function of static field for different high (low)
frequencies of the rf field from the Pt/Py/Al device
\cite{switch}. The dark (light) curves denote resonance at
$f_{low} (f_{high})$. (b) The static magnetic field
dependence of the resonance frequency of the Py strip
(dots). The curve is a fit to Eq. \ref{Kittel}. (c) The
amplitude of the dc voltage from a Al/Py/Al device as a
function of the square of the rf current, at 13 GHz and 139
mT (dots). The line shows a linear fit. \cite{mePRL}}
\label{fig3}
\end{figure}

\begin{figure}[h]
\centering
\includegraphics[width = 8cm]{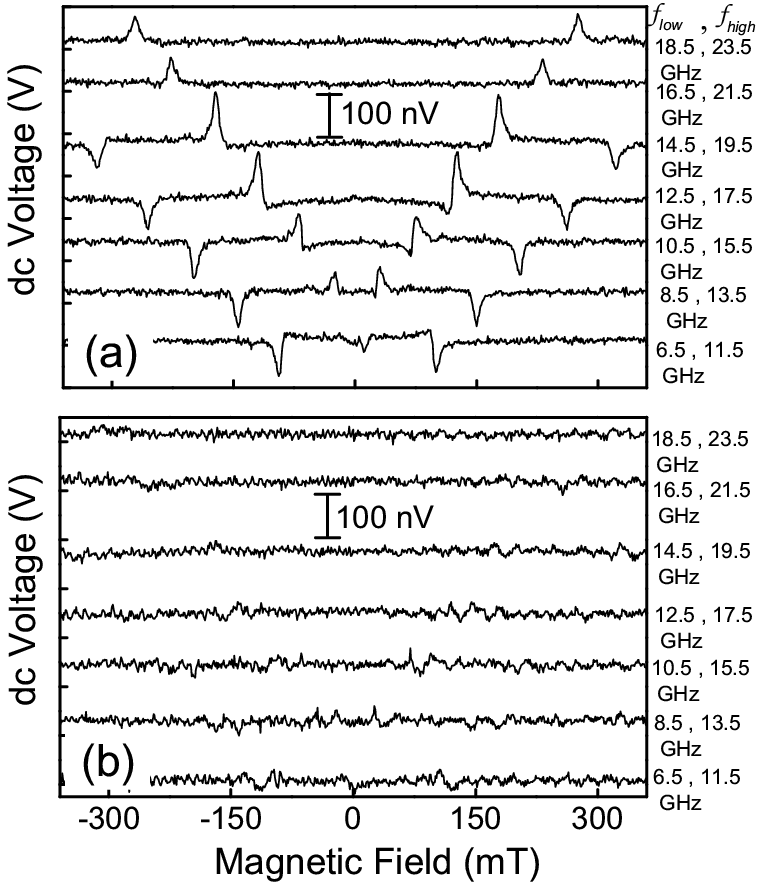}
\caption{The dc voltage $\Delta V$ generated across the Al/Py/Al (a) and Pt/Py/Pt (b) devices as a function of the static magnetic field. The frequencies of the rf field are as shown. \cite{mePRL}} \label{fig4}
\end{figure}
\begin{figure}[h]
\centering
\includegraphics[width = 6cm]{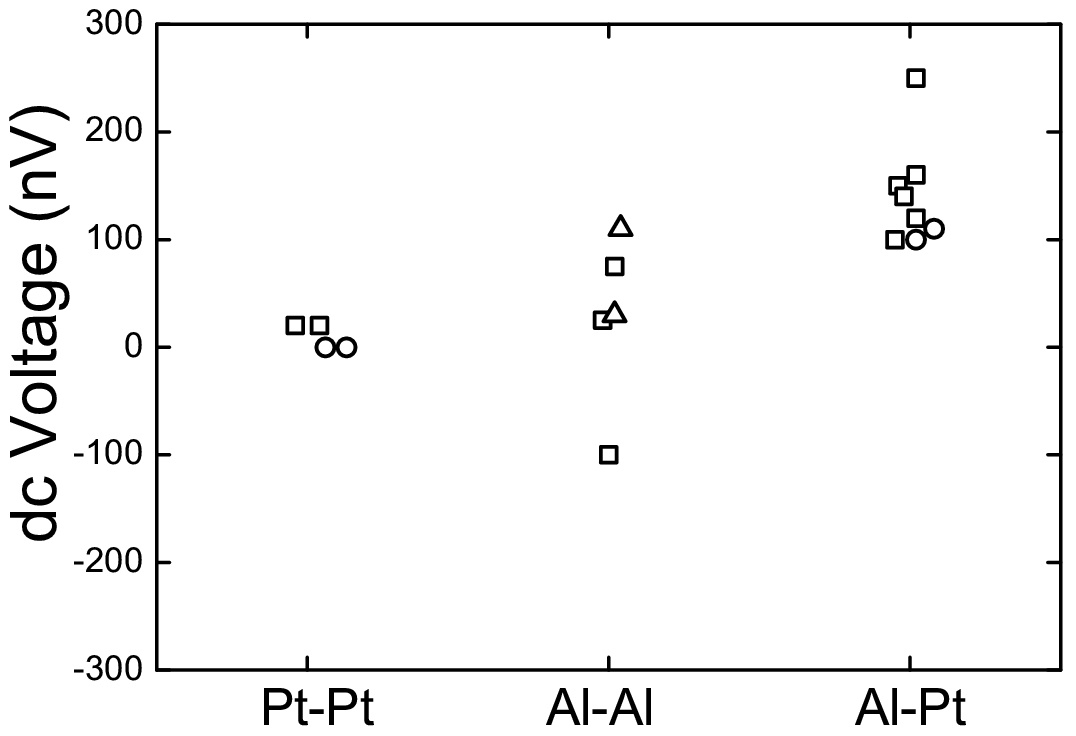}
\caption{Overall distribution of the amplitude of the dc voltages as a function of different contact materials. Different symbol represents different batch of samples. This includes 4 devices with longitudinal electrode device geometry, indicated by symbol ($\circ$) and discussed in section \ref{in-line}.}\label{distribution}
\end{figure}

First, we look at the peak/dip position dependence of the rf frequency. In Figure \ref{fig3}(a), the dc voltage in gray scale is plotted versus static field for different high (low) frequencies of the rf field. Figure \ref{fig3}(b) shows the fitting of the peak/dip position dependence of the rf field frequency (dotted curve) using Kittel's equation for a small angle precession of a thin-strip ferromagnet \cite{kittel}:
\begin{equation}
f=\frac{\gamma \mu_{0}}{2\pi}\sqrt{(H+N_{\parallel}M_{S})(H+N_{\perp}M_{S})}
\label{Kittel}
\end{equation}
where $\gamma$ is the gyromagnetic ratio, $N_{\parallel}$, $N_{\perp}$ are in-plane (along the width of the strip) and out-of-plane demagnetization factors and $M_{S}$ is the saturation magnetization. The fit to this equation (see Fig. \ref{fig3}(b)) gives $\gamma=176$ GHz/T, and $N_{\parallel}\mu_{0}M_{S}=60$ mT, $N_{\perp}\mu_{0}M_{S}=930$ mT, consistent with earlier reports \cite{grundler_apl,meACAMR}. The fit confirms that the dc voltage appears at the uniform ferromagnetic resonance mode of the Py strip. The measured amplitude of the dc voltage as a function of the square of the applied rf current, at $13$ GHz and 139 mT is shown in Fig. \ref{fig3}(c). Here, we observe a linear dependence on the square of the rf current, consistent with the prediction of the spin pumping theory, see Eqs. \ref{xuhui_ch2} and \ref{voltage_ch2}.

Further, we studied several control devices where both electrodes are of the same non-magnetic material, Al (5 devices) or Pt (4 devices). Here we expected no signal because of the reasons mentioned above. The results from Al/Py/Al devices show smaller signals than Pt/Py/Al devices, with a large scatter in amplitude and both with positive and negative sign for the resonance at $f_{high}$. Values for the 5 devices are -100 nV (shown in Fig. \ref{fig4}(a)), +25 nV, +30 nV, +75 nV and +110 nV. In contrast, all 4 Pt/Py/Pt devices exhibit only weak signals less than 20 nV,  with resonance signals barely visible, as in Fig. \ref{fig4}(b).

The overall values of the dc voltages as a function of different contact materials are shown in Fig. \ref{distribution}. We summarize the results as follow:
\\
First, the Pt/Py/Al devices have signals that are always positive, on average $150~nV$, and with a scatter comparable in amplitude to that of Al/Py/Al devices around zero. This scatter in the signal amplitude can be due to: (i) samples variation, due to different interface quality, not identical contacts (i.e. different overlap between the N electrodes and the Py strip, see Fig. \ref{str}(b)) and a small variation in distance between the Py strip and the CSW; (ii) different rf power at the end of CSW, due to different positions and contact resistance of the microwave probe on the CSW. These characteristics are difficult to estimate for each device.
\\
Second, we attribute the signals from Al/Py/Al devices to the asymmetry of the two contacts, possibly caused by small variation of the interfaces and in the contact geometry. Depending on the asymmetry, the signals therefore have a scatter around zero.
\\
Third, in the Pt/Py/Pt devices, independent of possible asymmetry, we expected and found very small signals. Therefore, we conclude that the signals measured with the Pt/Py/Al devices arise mainly from the Al/Py interface.

\subsection{Spin pumping vs. Rectification effects}\label{in-line}

\begin{figure}[h]
\centering
\includegraphics[width = 8.6cm]{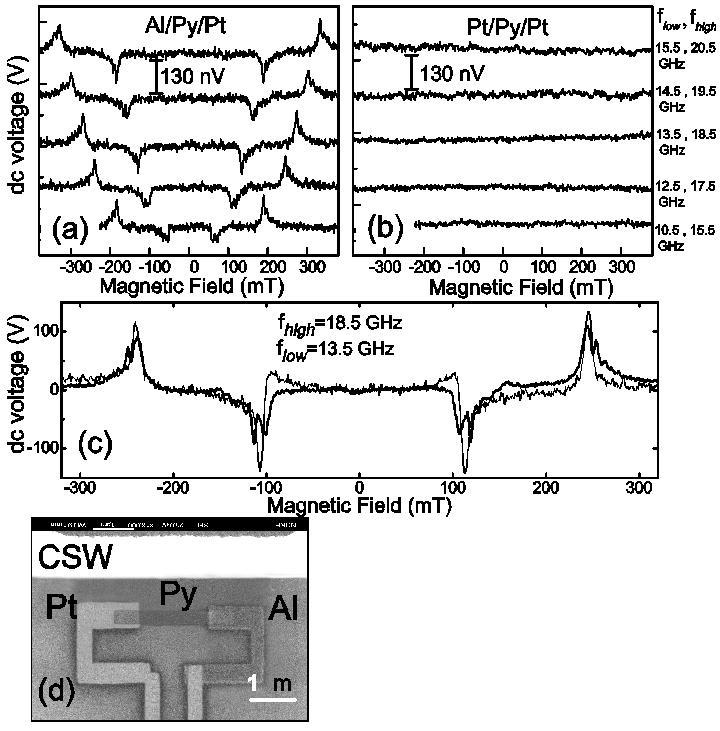}
\caption{The dc voltage generated by Al/Py/Pt (a) and
Pt/Py/Pt (b) devices for a device geometry shown in Fig.
(d). (c) The comparison of the signals for the two Pt/Py/Al
devices, one with longitudinal electrode device geometry
(bold line) and other with transverse electrode device
geometry. (d) SEM picture of an longitudinal electrode
geometry, Pt/Py/Al device.}\label{supliment1}
\end{figure}

We now discuss the rectification effects. As we have shown recently\cite{meAMR}, due to capacitive and inductive coupling between the CSW and the Py strip, rf currents ($I(t)=I_{0} \ cos \ \omega t$) can be induced in the detection circuit. The rf currents in combination with a time-dependent AMR ($R(t)\simeq \Delta R \ cos \ (\omega t+\varphi$)) can give a dc effect due to rectification effect ($V_{dc}\simeq \langle I\cdot R\rangle _{t}$). However, for rectification to occur, the resistance $R(t)$ must have first harmonic components, which is not true assuming circular or even elliptical precession of the magnetization.

There are two ways to have first harmonic components: (i) an offset angle between the applied field and the long axis of the Py strip, namely bulk rectification effect; (ii) an offset angle between the circulating rf currents and the magnetization. When the rf circulating currents enter and leave the strip, they can pass through a large angle relative to the magnetization. Asymmetry in the entry and exit paths, due to different conductivities of the two contacts, in combination with the time-dependent AMR, can lead to a rectification effect at the contacts, which we call the contact rectification effect \cite{me_thesis}.

Even if, we can accurately control the offset angle between the applied field and the Py strip, we cannot rule out the contacts effect that may also contribute to the data presented in the previous section, see Fig. \ref{fig2}. A small contribution from rectification effects on top of spin pumping signal can also explain the asymmetric peak/dip shape which does not have a Lorentzian shape as expected from Eq. \ref{xuhui_ch2}. In order to study these effects we prepared a new set of 4 devices very similar to the one shown in Fig. \ref{str}(b), but now with contacts at the ends of the Py strip, extending along the long axis of the strip, see Fig. \ref{supliment1}(d) for a SEM image. In this geometry, the induced rf current flows through the contacts predominantly parallel to the magnetization direction. This suppresses the possible contribution to the measured dc voltages from a rectification effect at the contacts \cite{rectif_prb}.

We first align the devices with Py strip parallel to the applied field and measure the dc voltage function of the field, as explained above. The measurements are shown in Fig. \ref{supliment1}(a),(b) for Pt/Py/Al and a Pt/Py/Pt configuration. These results are consistent with the above discussion, as the Pt/Py/Al devices show signals equal to the average value measured in the previous device geometry, while the Pt/Py/Pt devices show no signal as expected.
Figure \ref{supliment1}(c) shows a comparison between the voltages of two Al/Py/Pt devices with longitudinal (bold line) and transverse (normal line) contacts geometry, at $f_{high}$ = 18.5 GHz, $f_{low}$ = 13.5 GHz.
Particularly, devices with the longitudinal contacts exhibit, in addition to the main peak, a series of peaks at higher fields. An exact explanation of these observations is not yet clear. We assume these are related to end-mode resonances, since in this contacts geometry we are sensitive also to the magnetic structure of the ends of the Py strip. Moreover, we found no significant difference in the measured dc voltages between these two contacts geometries, Fig. \ref{str}(b) and Fig. \ref{supliment1}(d).
\begin{figure}
\includegraphics[width = 6.5cm]{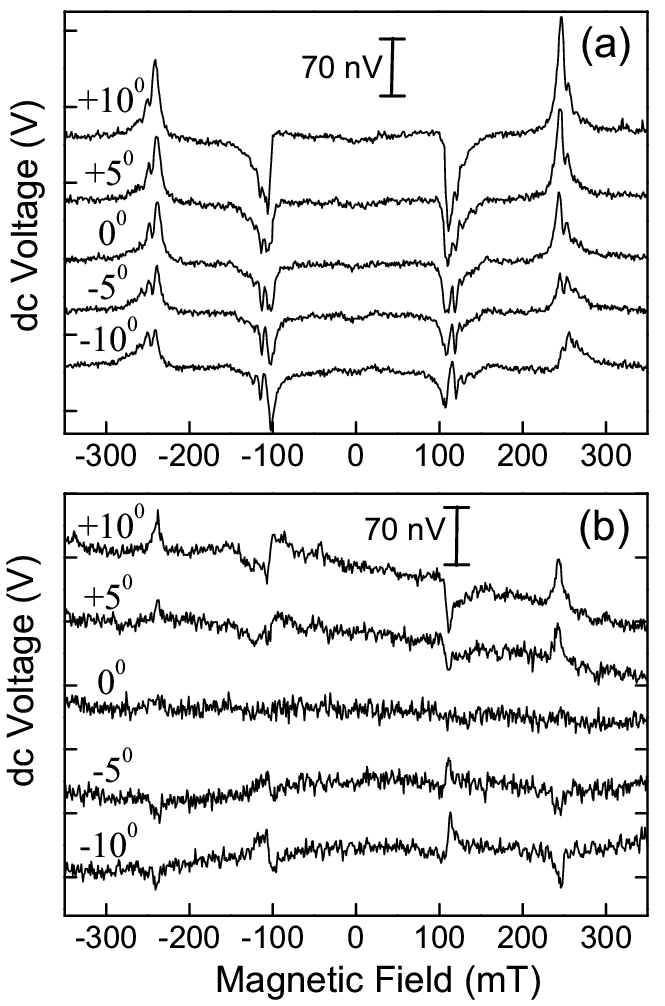}
\caption{The dc voltage at $f_{high}$ = 18 GHz, $f_{low}$ = 13 GHz for (a) Pt/Py/Al and (b) Pt/Py/Pt devices for different angles between the static field and the long axis of the strip.}\label{supliment2}
\end{figure}

\begin{figure}[h]
\centering
\includegraphics[width = 8.6cm]{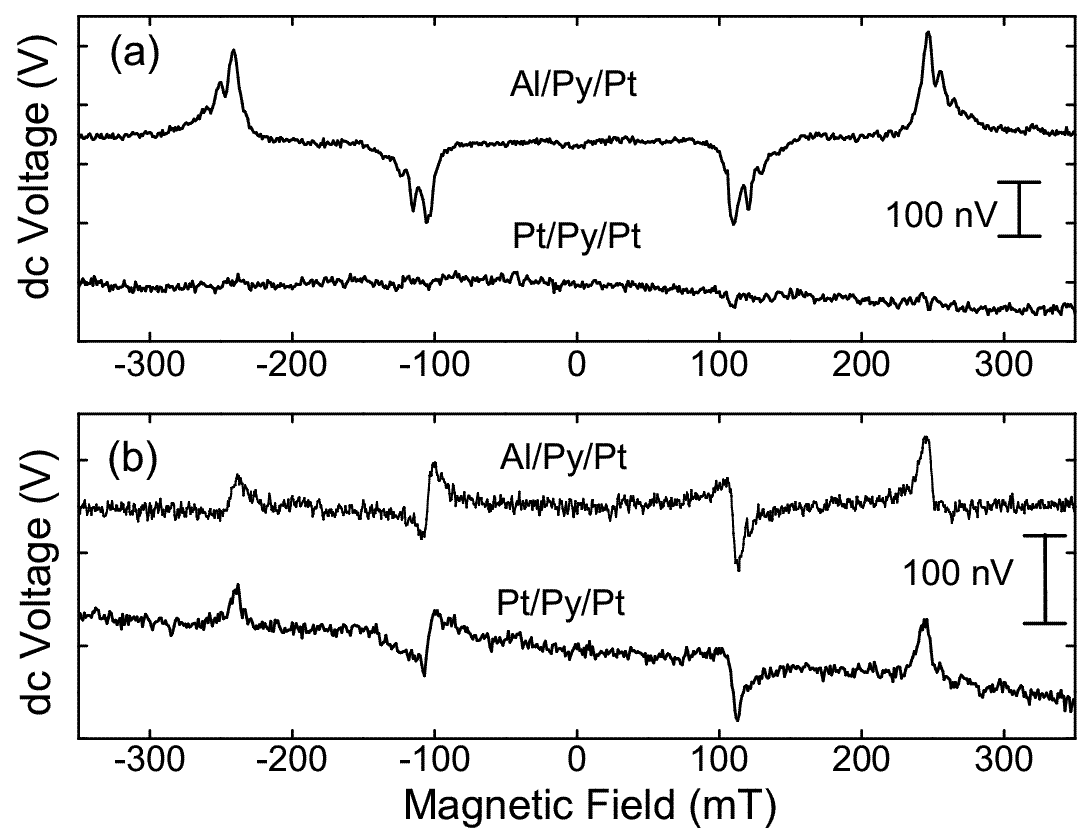}
\caption{To isolate the contribution from different effects we performed: (a) The sum between the voltage at offset angle of $10^{\circ}$ and $-10^{\circ}$,
$V(10^{\circ})+V(-10^{\circ})$, vs. static field for Pt/Py/Al and Pt/Py/Pt devices. The result represents 2 times contribution from spin pumping effect. (b) The difference between the voltages, $V(10^{\circ})-V(-10^{\circ})$, which represents 2 times contribution from bulk rectification effect, as explained in the text.}\label{supliment3}
\end{figure}

In order to confirm the above assumptions and to quantify the bulk rectification effect, we misaligned the direction of the static field with respect to the Py strip long axis by $5^{\circ}$ (and $10^{\circ}$) and measured the voltage at $f_{high}$ = 18 GHz, $f_{low}$ = 13 GHz. The results for Pt/Py/Al and Pt/Py/Pt devices are shown in Fig. \ref{supliment2}. Note that we see significant contributions from the bulk rectification effect only at offset angles larger than $5^{\circ}$. This rules out that small offset angles which may be present in the other geometry caused significant effects in the results, at most 10-20 nV.

In the following, the above results are analyzed taking into account that the voltages measured in Pt/Py/Al devices are due to two effects: (i) spin pumping, and (ii) bulk rectification effect for a non-zero offset angle. Of these two effects only the bulk rectification depends on the sign of the offset angle. This means that if we take the sum of the voltages measured at $+/-10^{\circ}$, $V(10^{\circ})+V(-10^{\circ})$, we obtain two times the contribution from the spin pumping effect with a Lorentzian peak shape. And in contrast we expect no signal if we do the same operation for Pt/Py/Pt devices. These results are shown in Fig. \ref{supliment3}(a).
\\
On the other hand, if we subtract, $V(10^{\circ})-V(-10^{\circ})$, we obtain two times the contribution from the bulk rectification effect. Figure \ref{supliment3}(b) shows the resulting data, which is practically the same for both devices, Pt/Py/Al and Pt/Py/Pt. Such a result is expected because the bulk rectification effect does not depend on the contact material.

We now consider a quantitative assessment of possible contribution to the measured
signal from rectification effects, namely bulk and contact rectification effect. Note, the contact rectification effect in principle should cancel for equivalent contacts. Both rectification effects depend primarily on the rf circulating current, which varies from device to device, depending on position of the pico-probe and rf current frequency. In a similar device geometry \cite{meAMR} we have estimated the rf currents, to be up to 30 $\mu$A. With this value we obtain:
\\
(i) Bulk: A rough estimate of an upper bound contribution, assuming an offset angle of 2$^{\circ}$, gives 15 nV. The data shown in Fig. \ref{supliment2}(b) (for zero degree) is less than this value.
\\
(ii) Contact: This contribution, which is present only in devices with the transverse electrode geometry, is estimated at 30 nV \cite{me_thesis}.
\\
The sum of these contributions can have any value between -45 and 45 nV, and thus can add or
subtract to the average spin pumping signal (150 nV), given the rise to extra scatter in the data, see Fig. \ref{distribution}.

In addition to signal magnitude analysis, it is also important to discuss the difference in signal shape due to these effects. It should be noted that each of the rectification effects discussed above can have a signal shape which can be any combination between absorptive and dispersive peak shape. In contrast, the spin pumping signal is only absorptive with a Lorentzian shape.

\section{Conclusion}

We have presented dc voltage due to the spin pumping effect, across the interface between Al and Py at ferromagnetic resonance. We found that the devices where the Al contact has been replaced by Pt show a voltage close to zero, in good agreement with theory. We observed a quadratic dependence of dc voltage function of precession cone angle, in agreement with the discussed theory. Theoretical predicted spin pumping voltage (20 nV) is less than the values observed experimentally (in average 150 nV). This underestimation might arise from the fact that the model does not consider device geometry, disorder at the interface and assumes an homogeneous magnetization in the ferromagnet.

Furthermore, to rule out a possible contribution from rectification effects to the measured signal, we have studied devices with different electrode geometries. We observed that for a non-zero offset angle, between the static field and the Py strip, the measured voltages are due to two different effects, namely spin pumping and rectification effects. By using an appropriate device geometry, these effects can be quantified and for the zero offset angle the rectification effects are minimized.

This work demonstrates a means of directly converting magnetization dynamics of a single nanomagnet into an electrical signal which can open new opportunities for technological applications.

\section*{ACKNOWLEDGMENTS}

We thank M. Sladkov, J. Grollier, A. Slachter, G. Visanescu, J. Jungmann for discussion and assistance in this project and B. Wolfs and S. Bakker for technical support.
This work was supported by the Dutch Foundation for Fundamental Research on Matter (FOM),
the Netherlands Organization for Scientific Research (NWO), the NanoNed and the DynaMax.

\bibliographystyle{apsrev}
\bibliography{../FJall}

\begin{thebibliography}{45}
\bibitem[{cur()}]{cur}
\bibinfo{note}{The 9 dBm applied rf power was converted into a rf current
  assuming $50~\Omega$ load impedance.}
\bibitem[{swi()}]{switch}
\bibinfo{note}{The irregular feature, only occurring near $H=0$ mT is due to
  magnetization switching with hysteresis.}
\bibitem[{rec()}]{rectif_prb}
\bibinfo{note}{Giving the arguments discussed, it is also posible a
  contribution from the rf currents flow in/out of plane. We have calculated
  this contribution and find that is negligible.}
\end{thebibliography}

\end{document}